\documentclass[a4paper, onecolumn, preprint]{revtex4}
\usepackage{amsmath}
\usepackage{amsfonts}
\usepackage{graphicx}
\usepackage{color}
\usepackage{lscape}
\usepackage{amssymb}

\begin{document}
\title{Charge-transfer state dynamics in all-polymer solar cells: Formation, dissociation and decoherence}

\author{Jiaqing Huang}
\affiliation{Department of Physics and State Key Laboratory of Luminescent Materials and Devices, South China University of Technology, Guangzhou 510640, China}

\author{Yijie Mo}
\affiliation{Department of Physics and State Key Laboratory of Luminescent Materials and Devices, South China University of Technology, Guangzhou 510640, China}

\author{Yao Yao}
\email{yaoyao2016@scut.edu.cn}
\affiliation{Department of Physics and State Key Laboratory of Luminescent Materials and Devices, South China University of Technology, Guangzhou 510640, China}

\begin{abstract}
All-polymer solar cells gained substantial achievements in recent years, offering numerous unsettled subjects for mechanical researchers. Based on the Su-Schrieffer-Heeger model, we then simulate the ultrafast dynamics of charge-transfer (CT) state considering a molecular electrostatic potential drop at the interface between two polymer chains, which are respectively regarded as donor and acceptor in all-polymer solar cells. The formation of a stable CT state is found to be sensitive to the distance between two oppositely charged polarons and the relevant critical electrostatic potential is thus quantified. In order to get insight into the dependence of dissociation of CT state on the width of interfacial layer, two quantities are calculated: One is the Coulomb capture radius between two polarons and the other is the quantum trace distance which serves as the fingerprint of the quantum coherence between them. The dissociation of CT state is found to take place within an ultrafast timescale for an optimum interfacial width. The classical spatial distance and the quantum trace distance manifest converging trend suggesting a decoherence scenario for the charge separation in all-polymer solar cells.
\end{abstract}

\maketitle

\section{INTRODUCTION}

Following the rapid developing progress of nonfullerene organic solar cells (OSCs) \cite{Zhan,Science,NM}, all-polymer solar cells (all-PSCs) emerged to be novel candidates with perfect performance and environmental friendliness \cite{Li1,Li2,Huang,Yao1,anie}. All-PSCs employ the conjugated polymers with strong electron-withdrawing ability such as the naphthalenediimide (NDI) polymer N2200 \cite{Li1,Ma,Zhou,Wang} as the electron acceptor which are qualitatively distinct from the conventional fullerene-based acceptors. They exhibit the advantages of tenability of electronic structure, enhanced light absorption, and superior mechanical and thermal properties \cite{Facchetti,Kang,Facchetti2}. The up-to-date power conversion efficiency (PCE) of all-PSCs has increased over $11\%$ \cite{PCE1,PCE2} benefitting from the optimized designment of polymeric materials for electron acceptor and the dramatic development of the device processing technology since all-PSCs were firstly reported in 1995 \cite{Halls}.

The study of the formation and efficient dissociation of the Frenkel excitons is crucial for understanding the working mechanisms of OSCs. The Frenkel exciton is conventionally regarded to be the initial local excited state after photoexcitation which possesses the characteristics of a tightly-bound electron-hole pair with large binding energy induced by both the self-trapping effect and the Coulomb attraction \cite{Friend,Chandross,Arkhipov}. It is desired that these excitons efficiently dissociate into free charge carriers in order to generate sufficiently large photocurrent. Nevertheless, an intermediate state composed of a weakly-bound electron-hole pair with the electron and the hole being respectively coupled to their own local lattice distortion, which is also called the charge-transfer (CT) state (or the polaron-pair state) \cite{Zhu1,Bredas,Zhu2,Gao,Sun}, is formed across the donor/acceptor (D/A) interface instead of the direct formation of free charge carriers. In traditional fullerene-based cells, the conversion from the Frenkel exciton to the CT state is determined by the relative relation between the exciton binding energy $\varepsilon_{\rm B}$ and the energy (ionization potential) offset ${\Delta}E$ between donor and acceptor \cite{Gao,Bittner1,Bittner2,Rand}. If ${\Delta}E$ is greater than $\varepsilon_{\rm B}$, the photogenerated Frenkel exciton will be transited into CT states owing to the energy instability; otherwise, the Frenkel exciton will keep stable and the dissociation will unlikely take place.

In nonfullerene solar cells, however, experiments have shown that the energy offset does not matter in the process of charge separation \cite{Liu,Nikolis}. Unlike the traditional fullerene-based cells in which a significant energy offset ($\sim0.3 {\rm eV}$) \cite{Arkhipov,Bittner1,Bittner2,Sebastian} is necessary for the charge separation, nonfullerene cells enable fast and efficient charge separation despite of the negligible energy offset ($\sim0.05 {\rm eV}$). Liu \textit{et al.} reported a nonfullerene OSC with PCE being $9.5\%$ based on the P3TEA:SF-PD${\rm I_2}$ blend \cite{Liu}, and Nikolis \textit{et al.} obtained high external quantum efficiency (EQE) of $79\%$ based on the $\alpha$-6T/SubNc nonfullerene OSC \cite{Nikolis}. Both their devices are observed to have a small driving force, which results in the low open-circuit voltage loss and thus improves the device efficiency. In this context, the current interest of the emerging nonfullerene OSCs is focusing on the following questions: Is there still an intermediate state such as the CT state in the process of charge separation, and what then serves as the driving force therein?

The dissociation of CT state has been widely investigated for a long history and many experimental and theoretical results of this process have been reported \cite{Deibel1,Rubel,Deibel2,Gregg,F. Gao,Zhu3,Ono,Hou1,Gundogdu,Han,Hou2}. The energetic disorder was a crucial point in the early investigations, and based on the hopping transport model the field-assisted dissociation of CT state was proposed in the disordered systems \cite{Rubel}. The surface losses due to CT state diffusion were taken into account for the dissociation efficiency of CT state \cite{Deibel1}. By comparison, the critical effects of charge delocalization and entropy increase were highlighted in the process of CT state dissociation \cite{Deibel2,Gregg,F. Gao,Zhu3,Ono}. More recently, the molecular packing, orientation and blend morphology were also reported to play important roles in CT state dissociation and charge separation \cite{Kang,Hou1,Gundogdu,Han}. Despite of these researches, the underlying mechanism of the dissociation of CT state is still hotly debated. Jianhui Hou's group recently proposed a novel mechanism based upon the molecular electrostatic potential (ESP) \cite{Hou2}. They stated that the intermolecular electric field resulting from the difference of the molecular ESPs between donor and acceptor materials facilitated efficient charge separation at the D/A interface in nonfullerene cells.

In recent works, one of the authors studied the charge transfer and separation in small-molecule OSCs, in which the long-range charge transfer state was highlighted \cite{Yao2,Yao3}. On the other side, herein, the discussion of all-PSCs is presented. Two components of the polymer chain are set head-to-tail to construct the D/A interface and a molecular ESP drop is involved in the modeling. The formation and the dynamical dissociation of CT state at the D/A interface are investigated within the framework of Su-Schrieffer-Heeger (SSH) model \cite{Su}. The conditions for the existence of a stable CT state will be discussed, and the separation process of the two polarons in the CT state will be featured by the spatial distance, as well as the quantum trace distance. The paper is organized as the following sequence. The model and method are given in Section \uppercase\expandafter{\romannumeral2}. The results and discussion are presented in Section \uppercase\expandafter{\romannumeral3}. Finally in Section \uppercase\expandafter{\romannumeral4} the main conclusions are drawn.

\section{MODEL AND METHOD}

In a long history, the polythiophene (PT) is the most commonly-used electron donor in PSCs. Electron acceptors in all-PSCs, such as that in the blend of J51/N2200 introduced by Li's group \cite{Li1}, also adopt thiophenes as core groups to polymerize NDIs. Radicals on the imide end groups in NDIs have got strong electron affinity, and additional carriers are thus self-doped into the backbone formed by thiophenes. The alkyl groups behave as side groups to improve solubility and form a perfect one-dimensional (1D) polymer chain. All these features allow us to employ the benchmarking SSH model to mimic the microscopic physics of all-PSCs. The representative PT chain is investigated as a model system to study the dynamics of CT state in the all-PSCs. The $\pi$-conjugated orbits of the thiophene backbone provide the transport sites for the self-doped electrons from imide groups and other radicals. In order to study the D/A interface, two polymer chains with different electron affinity (molecular ESP) are placed head-to-tail.

\begin{figure}
\centering
\includegraphics[scale=0.75]{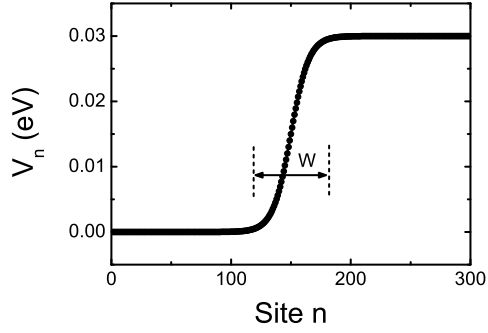}
\caption{The spatial distribution of the on-site energy $V_{n}$. $W$ is the width of the interfacial layer.
}\label{figure1}
\end{figure}

The total model Hamiltonian then consists of three terms which has the following form:
\begin{equation}\label{total hamiltonian}
H=H_{\rm SSH}+H_{\rm U}+H_{\rm P}.
\end{equation}
Herein, the first term represents the original SSH Hamiltonian which contains two parts,
\begin{equation}\label{SSH Hamiltonian}
H_{\rm SSH}=H_{\rm ele}+H_{\rm lat}.
\end{equation}
$H_{\rm ele}$ in Eq.~(\ref{SSH Hamiltonian}) is the Hamiltonian of electrons expressed as
\begin{equation}\label{electronic hamiltonian}
H_{\rm ele}=-\sum_{n}t_{n} (\hat{c}_{n+1}^{\dagger} \hat{c}_{n}+{\rm h.c.}),
\end{equation}
where $\hat{c}_{n}^{\dagger}$ ($\hat{c}_{n}$) creates (annihilates) an electron on $n$-th site and the electron-phonon (e-p) interaction is involved in the nearest-neighbor hopping integral $t_{n}$ given by
\begin{equation}\label{hopping integral}
t_{n}=t_{0}-\alpha(u_{n+1}-u_{n}),
\end{equation}
with $t_{0}$ being the hopping constant, $\alpha$ the e-p coupling strength and $u_{n}$ the displacement of $n$-th thiophene unit. The second term in Eq.~(\ref{SSH Hamiltonian}) represents the elastic potential and kinetic energy of thiophene unit on the backbone, that is
\begin{equation}\label{lattice hamiltonian}
H_{\rm lat}=\frac{K}{2}\sum_{n} (u_{n+1}-u_{n})^2+\frac{M}{2}\sum_{n} \dot{u}_{n}^2,
\end{equation}
with $K$ being the elastic constant and $M$ being the mass of thiophene unit. The second term of Eq.~(\ref{total hamiltonian}) is for the many-body electron couplings, which quantifies the coupling between electron and hole, and the form is
\begin{equation}\label{e-e interaction hamiltonian}
H_{\rm U}=U\sum_{n} \hat{c}_{n,\uparrow}^{\dagger}\hat{c}_{n,\uparrow}\hat{c}_{n,\downarrow}^{\dagger}\hat{c}_{n,\downarrow},
\end{equation}
where $U$ gives the strength of the Coulomb interactions. This term will be treated with the Hartree-Fock approximation in the calculations. As the critical consideration of this work, the third term of Eq.~(\ref{total hamiltonian}) is the on-site energy $V_{n}$ denoting the different electron affinities (molecular ESPs) of donor and acceptor, i.e.,
\begin{equation}\label{on-site energy hamiltonian}
H_{\rm P}=\sum_{n} V_{n} \hat{c}_{n}^{\dagger}\hat{c}_{n}.
\end{equation}
In order to make the change of $V_{n}$ across the interface smooth, an analytic form is set as
\begin{equation}\label{on-site energy}
V_{n}=\frac{V_{0}}{2} \left[\tanh{\frac{4a_0(n-n_{0})}{W}}+1\right],
\end{equation}
where $V_0$ is the potential drop between donor and acceptor, $n_{0}$ is set to the central site of the chain, $W$ is the width of the interfacial layer and $a_0$ is the lattice constant. Fig.~\ref{figure1} displays the spatial distribution of $V_{n}$ with the total site number being 300. The donor component of the chain is labeled from 1 to 150 and the remaining is appointed to be the acceptor component. There is an interfacial layer between these two components that we can call it as the D/A interface. $V_{0}$ and ${W}$ determine the electric field strength induced by the potential drop, and obviously at the center of the interface the electric field is the strongest. It is noted that, the analytic form in Eq.~(\ref{on-site energy}) is not necessary to be the hyperbolic tangent function and the results in the following are not sensitive to this specific choice.

The time evolution of the electronic wave function is described with the time-dependent Schr\"odinger equation:
\begin{equation}\label{S equation}
i\hbar\frac{\partial{\psi_{\mu,n}(t)}}{\partial{t}}=-t_{n} \psi_{\mu,n+1} (t)-t_{n-1} \psi_{\mu,n-1} (t),
\end{equation}
with $\psi_{\mu,n}$ being the $\mu$-th eigen-state of the Hamiltonian (\ref{total hamiltonian}) on $n$-th site. The solution of this time-dependent Schr\"odinger equation is formally written as
\begin{equation}\label{solution}
\psi_{\mu}(t)=\hat{T} \exp\left[-\frac{i}{\hbar} \int_{0}^{t}{\rm d}{t'} H ({t'})\right] \psi_{\mu}(0),
\end{equation}
where $\hat{T}$ denotes the time ordering operator. In order to obtain the numerical solution of the electronic wavefunction, the integrate time step ${\Delta}t$ must be set to sufficiently small, namely below the order of the bare phonon frequency $\omega_{\rm Q}=\sqrt{4K/M}$. Throughout this work, we set it to $0.2 {\rm fs}$. The equation is then rewritten as
\begin{equation}\label{solution}
\psi_{\mu}(t_{j+1})=\exp\left[-iH(t_{j}){\Delta}t/{\hbar}\right] \psi_{\mu}(t_{j}).
\end{equation}
This equation can be alternatively expressed by the instantaneous eigenfunctions $\varphi_{\nu}$ and eigenvalues $\varepsilon_{\nu}$ of the Hamiltonian $H(t_{j})$, i.e.,
\begin{equation}\label{solution}
\psi_{\mu}(t_{j+1})=\sum_{\nu} \left\langle\varphi_{\nu}\mid{\psi_{\mu} (t_{j})}\right\rangle \exp\left[-i\varepsilon_{\nu} {\Delta}t/{\hbar}\right] \varphi_{\nu}.
\end{equation}

The motion of the lattice site is described as:
\begin{equation}\label{force}
F_{n} (t)=M\ddot{u}_{n}=-K[2u_{n}-u_{n+1}-u_{n-1}]+\alpha[\rho_{n,n+1}-\rho_{n-1,n}+\rho_{n+1,n}-\rho_{n,n-1}],
\end{equation}
where $F_{n} (t)$ is the force exerted on the ${n}$-th site and the density matrix $\rho$ is given by
\begin{equation}\label{density matrix}
\rho_{n,{n'}}=\sum_{\mu}\psi_{\mu,n}^{\ast} (t)f_{\mu} \psi_{\mu,n'} (t),
\end{equation}
with $f_{\mu}$ being the time-independent distribution function, which equals to 0, 1 or 2 and reflects the occupation of the electrons on the energy levels of the single-partite system. The lattice displacement $u_{n} (t_{j+1})$ and the velocity $\dot{u}_{n} (t_{j+1})$ can then be obtained with the following forms:
\begin{equation}\label{lattice displacement}
u_{n}(t_{j+1})=u_{n} (t_j)+\dot{u}_{n}(t_j){\Delta}t,
\end{equation}
\begin{equation}\label{velocity}
\dot{u}_{n}(t_{j+1})=\dot{u}_{n}(t_j)+{\frac{F_{n} (t_j)}{M}} {\Delta}t.
\end{equation}

In this work, we do not study the instantaneous processes of photoexcitation, emission and conversion from exciton to CT state. The dynamics we consider merely takes place on one potential surface, so that the conical cross of surfaces for different states does not matter. It is also worth noting that when the conversion processes between states are investigated, the conical cross plays an irreducible role, which is held as our future subject.

During the dynamical dissociation process of CT state, the description of the motion of the two oppositely charged polarons requires us to define the charge center of a polaron, which is of the following form:
\begin{equation}\label{charge centers}
x_{\rm c}=
\begin{cases}
L{\theta}/{2\pi},       & \text{if ${\left\langle\cos{\theta_n}\right\rangle}\geq{0}$ and ${\left\langle\sin{\theta_n}\right\rangle}\geq{0}$} \\
L({\theta}+\pi)/{2\pi},   & \text{if ${\left\langle\cos{\theta_n}\right\rangle}\leq{0}$}  \\
L({\theta}+2\pi)/{2\pi},  & \text{otherwise}
\end{cases}
\end{equation}
where
\begin{equation}\label{cos}
\left\langle\cos{\theta_n}\right\rangle=\sum_{n}\rho_{n}\cos(2{\pi}n/L),
\end{equation}
\begin{equation}\label{sin}
\left\langle\sin{\theta_n}\right\rangle=\sum_{n}\rho_{n}\sin(2{\pi}n/L),
\end{equation}
\begin{equation}\label{theta}
\theta=\arctan\left(\frac{\left\langle\sin{\theta_n}\right\rangle}{\left\langle\cos{\theta_n}\right\rangle}\right)
\end{equation}
and the net charge density is $\rho_{n}=\rho_{n,n}-1$. In order to distinguish the sign of polarons, the site index ${n}$ runs over the sites of donor and acceptor, respectively. The parameter $L$ represents the lattice number of each component.

If the two polarons in the CT state and also the two separated ones are regarded as two different partite, how do we identify the quantum correlations between these two quantum states? The trace distance is one of the extensively used measures that distinguish two quantum states accurately, which can be expressed as \cite{Fuchs,Breuer,Aaronson}
\begin{equation}\label{trace distance}
D(\rho,\Omega)=\frac{1}{2}{\rm Tr}\sqrt{(\rho-\Omega)^{\dagger} (\rho-\Omega)},
\end{equation}
where $\Omega$ denotes the density matrix of the initial CT state and $\rho$ represents the instantaneous density matrix of the system at time $t$.

\section{RESULTS AND DISCUSSION}

We first discuss the formation of the CT state. In the practical simulations, we first calculate the ground state of the entire system, and then a $\pi$-electron is excited from the highest occupied molecular orbit (HOMO) to the lowest unoccupied molecular orbit (LUMO), which mimics the process resulting from the photoexcitation in the realistic condition. Via this manipulation, we can obtain either a Frenkel exciton or a CT state. In order to focus on the latter, we artificially put two self-trapping potential valleys (lattice distortions) in the system to induce two oppositely charged polarons, one of which is put in donor and the other in acceptor. The initial distance between two valleys could be realized as the {\it size} of the CT state. Through the subsequent procedure of energy optimization for the entire system, the two polarons keep stable in some cases as discussed below, so that we can say the CT state robustly constitutes the excited state after photoexcitation, otherwise the CT state does not emerge as a photoexcited state. Furthermore, in order to display a good visualization of the dynamical results, we calculate the smoothed form of the lattice configuration for the displacement of each site and net charge density in the following:
\begin{equation}\label{smoothed lattice configuration}
\tilde{u}_n (t)=(-1)^n \left[2u_n (t)-u_{n-1} (t)-u_{n+1} (t)\right]/4,
\end{equation}
\begin{equation}\label{smoothed net charge density}
\tilde{\rho}_n (t)=\left[2\rho_n (t)+\rho_{n-1} (t)+\rho_{n+1} (t)\right]/4.
\end{equation}
In all the simulations, we take J51/N2200 as example to investigate and the values of parameters are set as follows \cite{para}: $t_{0}=0.09 {\rm eV}$, $\alpha=3.4 {\rm eV/\r{A}}$, $K=231 {\rm eV/\r{A}^2}$, $M=8509.96 {\rm eV\cdot fs/\r{A}^2}$ and $a_0=3.9 {\rm \r{A}}$.

\begin{figure}
\centering
\includegraphics[scale=0.75]{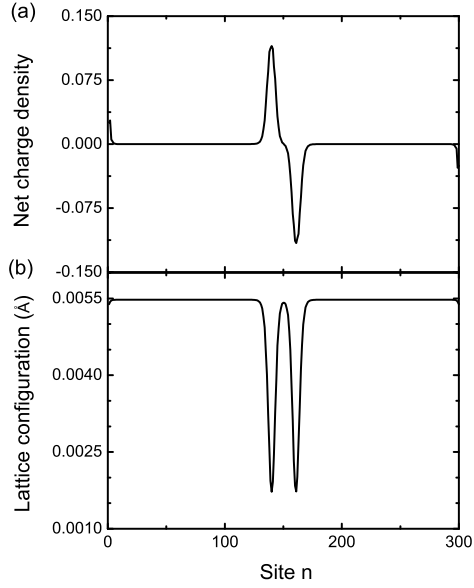}
\caption{(a) The net charge density  of CT state and (b) the lattice configuration for the site displacements with $W=60a_0$, $V_0=0.03 {\rm eV}$, $U=0.02 {\rm eV}$ and $a_0$ being the lattice constant.
}\label{figure2}
\end{figure}

\begin{figure}
\centering
\includegraphics[scale=0.75]{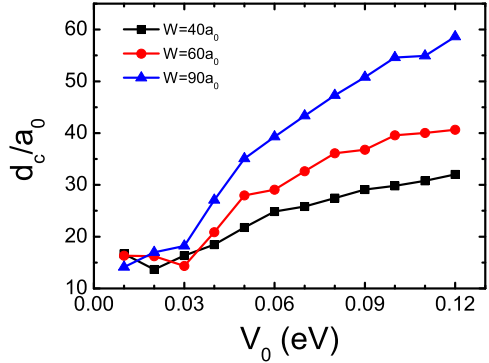}
\caption{The critical value of distance $d_{\rm c}$ between two oppositely charged polarons versus the molecular ESP drop $V_0$ for three interfacial widths with $U=0.02 {\rm eV}$.
}\label{figure3}
\end{figure}

\begin{figure}
\centering
\includegraphics[scale=0.75]{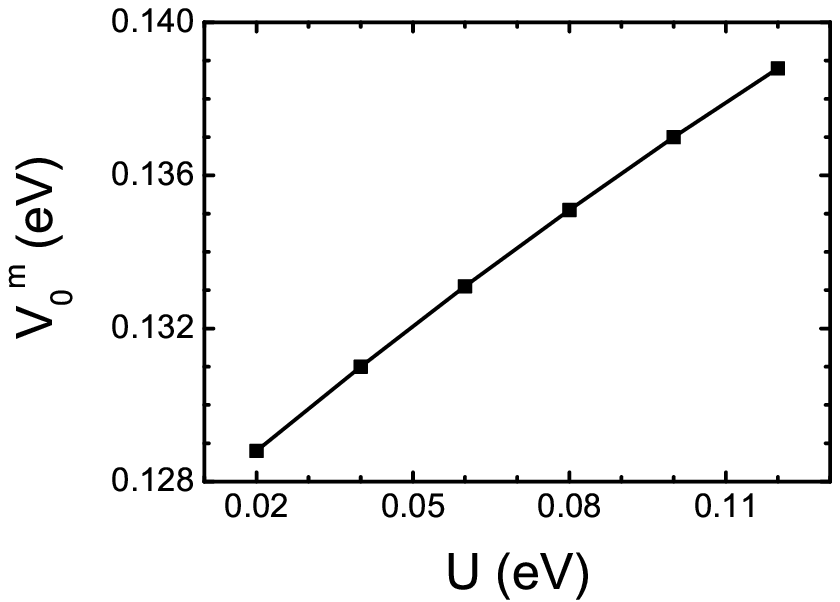}
\caption{The maximum value of molecular ESP drop $V_0^{\rm m}$ for the formation of stable CT state versus the Coulomb interaction $U$ with $W=60a_0$.
}\label{figure4}
\end{figure}

As an example of the emergence of CT state, we show in Fig.~\ref{figure2} for the case of $W=60a_0$, $V_0=0.03 {\rm eV}$ and $U=0.02 {\rm eV}$. Two localized states of electrons in a CT state are observed clearly, which correspond to two oppositely charged polarons respectively as shown in Fig.~\ref{figure2}(a). As stated above, these two localized electronic states can induce two local lattice distortions on the molecular chain, which is presented in Fig.~\ref{figure2}(b). These two polarons interact with each other via both the self-trapping valleys and the Coulomb interaction, and the shorter the distance $d$ between them is, the stronger the effective binding energy is. Here, the distance $d$ is quantified by the difference of charge centers between two polarons, i.e., $d={\Delta}x_{\rm c}$. When $d$ is shorter than a critical value $d_{\rm c}$, the attractive interaction in between would pull the two polarons together and merge them into a single self-trapping valley to form a Frenkel exciton, and in this situation the CT state can not be robustly formed. Therefore, the emergence of a robust CT state naturally refers to the value of $d_{\rm c}$ which strongly depends on the interfacial width $W$, the molecular ESP drop $V_0$ and the Coulomb interaction $U$. In addition, the molecular ESP drop $V_0$ acts as a driving energy to dissociate the CT state, and when $V_0$ is too strong the polaron itself is not stable, so is the CT state.

Fig.~\ref{figure3} displays $d_{\rm c}$ at different $V_0$ for three values of interfacial width $W$ with $U=0.02 {\rm eV}$. There are small fluctuations stemming from numerical errors since the driving force is very flat in a large extent of parameters. It is exhibited that the narrower interfacial layer is more helpful to facilitate the formation of robust CT state with the shorter distance $d_{\rm c}$. Moreover, when $V_0$ is smaller than $0.03 {\rm eV}$ the critical distance $d_{\rm c}$ changes slightly indicating the binding and the driving energy are easy to be balanced in this case. As $V_0$ increases from $0.03 {\rm eV}$ to $0.12 {\rm eV}$, $d_{\rm c}$ increases as well to weaken the electric field and balance the binding energy. It is also found that when $V_0$ is larger than $0.13 {\rm eV}$ the system cannot spontaneously form two localized polarons no matter how wide the interface is, suggesting the ESP drop is too large to induce a stable polaron state. We also observe that the maximum value of molecular ESP drop $V_0^{\rm m}$ for the emergence of CT state is sensitive to the Coulomb interaction $U$: The stronger the interaction is, the larger the $V_0^{\rm m}$ is. Fig.~\ref{figure4} shows that $V_0^{\rm m}$ increases almost linearly with $U$ increasing further corroborating the scenario above. Consequently, the question whether a CT state is formed in all-PSCs is parametrized in our model which can be examined in experimental researches.

\begin{figure}
\centering
\includegraphics[scale=0.75]{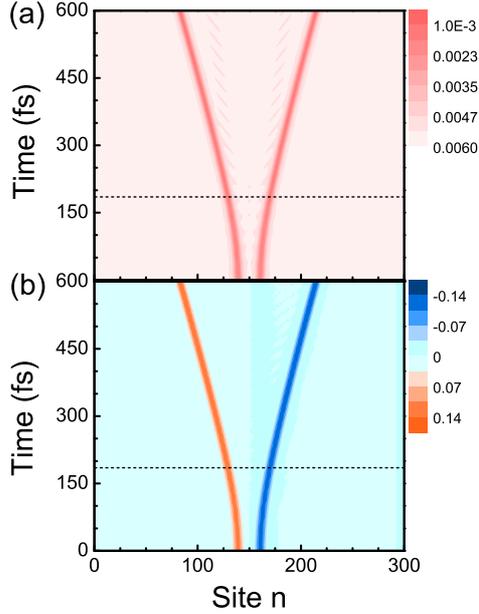}
\caption{Time evolution of (a) the lattice configuration $\tilde{u}_n (t)$ and (b) the net charge density $\tilde{\rho}_n (t)$ for CT state dissociation at the D/A interface with $W=60a_0$, $V_0=0.03 {\rm eV}$ and $U=0.02 {\rm eV}$. The dashed lines indicate the dissociation time of CT state.
}\label{figure5}
\end{figure}

The dynamics of the CT state are simulated via the nonadiabatic dynamical method at different interfacial widths from 10$a_0$ to 120$a_0$ with $V_0=0.03 {\rm eV}$, $U=0.02 {\rm eV}$. The initial distance between the two oppositely charged polarons are set to be around $d_{\rm c}$, namely, the minimum size of the CT state. In details, the two polarons initially reside on site 142 and 158, respectively, so the initial size of the CT state is set to 16$a_0$. As displayed in Fig.~\ref{figure5}, following time evolving, the two polarons move separately along the chain enabled by the driving force provided by the ESP drop at the D/A interface. With the increasing of distance between them, the Coulomb attraction of these two polarons tends to be weak until it can be ignored when the two polarons are far enough away from each other. The current interest is then, how to determine the dissociation of the CT state giving the dynamics of the two oppositely charged polarons. According to the Onsager theory \cite{Onsager}, when the Coulomb attraction between two polarons is equal to or smaller than the thermal energy $k_{\rm B}T$ at room temperature, the two polarons can be regarded to dissociate, namely the CT state dissociate into free polarons giving rise to the generation of photocurrent. The distance at which the two polarons are regarded to be free is the so-called Coulomb capture radius defined as
\begin{equation}\label{capture radius}
R_{\rm c}=\frac{e^2}{4\pi{\varepsilon_{\rm r}} \varepsilon_{0} k_{\rm B} T},
\end{equation}
where $e$ is the elementary charge of electron, $\varepsilon_{\rm r}$ is the relative dielectric constant, $\varepsilon_{0}$ is the permittivity of vacuum, $k_{\rm B}$ is the Boltzmann's constant and $T$ is temperature. For most organic molecules and polymers, the dielectric constant ranges between 3 and 4 \cite{Arkhipov}, so we can simply set it to 3.5. The Coulomb capture radius $R_{\rm c}$ is then calculated to be around $160 {\rm \r{A}}$ (i.e., $\sim$40$a_0$) in our case. By this definition, we can find that the dissociation time $T_{\rm D}$ of CT state, defined as the time point that the distance between two polarons equals to $R_{\rm c}$, is obtained to be $\sim190 {\rm fs}$ denoted by the dashed line in Fig.~\ref{figure5} with the interfacial width being 60$a_0$, suggesting that the CT state dissociation takes place in an ultrafast timescale.

\begin{figure}
\centering
\includegraphics[scale=0.75]{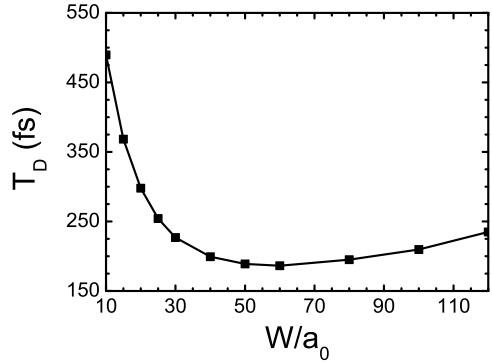}
\caption{The dissociation time $T_{\rm D}$ of CT state (see the text for the definition) versus the interfacial width.
}\label{figure6}
\end{figure}

A suitable interface for efficient dissociation in the D/A heterojunction is crucial to the efficiency of OSCs. The experiments have figured out that the dissociation of CT state is sensitive to the scale of the phase separation of D/A blend films in nonfullerene cells \cite{McNeill,Bao,Melkonyan}. It is thus interesting to make a comparison of the dissociation time $T_{\rm D}$ among various interfacial widths, which is shown in Fig.~\ref{figure6}. With increasing of the interfacial width, the dissociation time dramatically decreases when $W<60a_0$ followed by a slow increase. When the interface has a relatively small width of 10$a_0$, the CT state takes $\sim 490 {\rm fs}$ to dissociate. This is because in the initial state the two polarons are partly located outside the interfacial region and do not feel the driving force induced by the ESP drop very much. It is obviously that the dissociation of CT state inside the interfacial region is much more efficient than that outside the interface. When the interfacial width is larger than 60$a_0$, the two polarons are completely residing in the interfacial region, but the electric field and thus the driving force induced by the ESP drop becomes smaller leading to the slow dissociation of CT state. Consequently, the optimum value of the interfacial width for the efficient dissociation time is determined to be 60$a_0$ ($\sim23 {\rm nm}$) in our case. Despite of the quantitative difference in the realistic solar cells, the interfacial width of around 60$a_0$ dominated by the phase separation of two polymer chains in the heterojunction structure is the most favorable for the dissociation of CT state, in good agreement with the experimental results \cite{Li2,Yan}.

\begin{figure}
\centering
\includegraphics[scale=0.75]{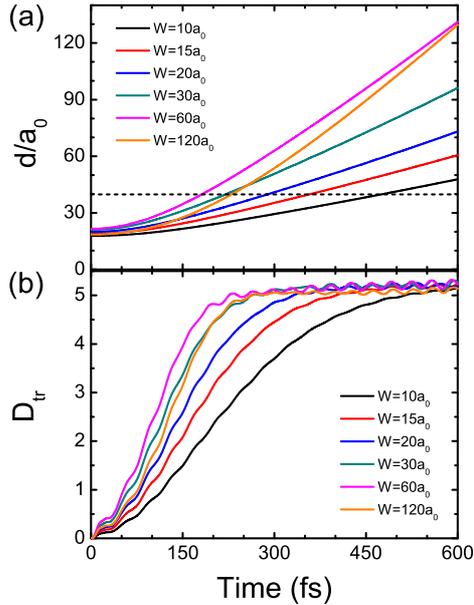}
\caption{Time evolution of (a) the spatial distance $d$ between two polarons and (b) the trace distance $D_{\rm tr}$ for CT state dissociation various interfacial widths. The dashed line reflects the Coulomb capture radius $R_{\rm c}$.
}\label{figure7}
\end{figure}

One would be doubting that when we are discussing an ultrafast process for the dissociation of CT state, the introduction of Coulomb capture radius based on the thermal fluctuation at room temperature does not make sense. It is thus contributive to give a more enlightening measure for the dissociation based upon the quantum dynamics. The quantum trace distance, defined in Eq.~(\ref{trace distance}), is introduced to quantify the dissociation as shown in Fig.~\ref{figure7}. As a comparison, Fig.~\ref{figure7}(a) first displays the time evolution of the spatial distance between two polarons. With time increasing, the spatial distance $d$ increase approximately linearly for all interfacial widths and $W=60a_0$ is the optimum value for the efficient dissociation of CT state, as discussed above. Fig.~\ref{figure7}(b) exhibits the time evolution of the trace distance in the dissociation process of CT state. Despite of the difference of interfacial width, the trace distance $D_{\rm tr}$ behaves similarly. Namely, it firstly increases with time increasing and afterward remains approximately constant suggesting the quantum coherence between the two polarons vanishes. It is found that, the time point at which the trace distance gets to be constant, namely the two polarons completely lose the coherence, exhibits significant difference for various interfacial widths. It is the fastest to completely lose the coherence when the interfacial width is 60$a_0$. Below this value, the time gets significantly longer. The narrower the interface is, the longer the time is. Comparing with Fig.~\ref{figure7}(a), we can find that the time of the dissociation of CT state based upon the Coulomb capture radius is coincidentally close to that for losing the coherence at different interfacial widths. Along with the ultrafast timescale, this implies that the dissociation of the CT state manifests a decoherence scenario, and in a quantitative manner the classical distance between polarons can serve as a featured measure of the dissociation, accompanying with the quantum trace distance.

\section{CONCLUSIONS}

In this work, we theoretically investigate the formation and dynamical dissociation of CT state at the D/A interface. We modify the extensively used SSH model with a molecular ESP drop and perform the simulations with the nonadiabatic dynamical method. It is found that the formation of CT state depends on the width of the interfacial layer, the ESP drop between the interfacial layer and the Coulomb interaction. A maximum ESP drop for the robust formation of CT state is observed. The dynamical dissociation process of CT state is then discussed. With the Coulomb capture radius being the criterion, it is obtained that the two oppositely charged polarons in the CT state separate completely at the D/A interface within hundreds of femtosecond. The D/A interface of $23 {\rm nm}$ is the most suitable interfacial layer for the dissociation of CT state. In addition, the quantum trace distance provides the further demonstration on the dissociation of CT state, which can explain the physical meaning of charge separation properly at the D/A interface. Our work can be of practical significance to the optimization of the all-PSCs.

\section*{Acknowledgment}

The authors gratefully acknowledge support from the National Natural Science Foundation of China (Grant Nos.~11574052 and 91333202).


\begin{thebibliography}{60}
\bibitem{para}  The hopping integrals $t_0$ for both J51 and N2200 are calculated to be around 0.09eV by our quantum chemistry computations. The parameters for the lattice part can be determined by considering the relevant vibrational spectra. For N2200, e.g., the primary bare optical phonon mode $\omega_{\rm Q}$ is 1769 cm$^{-1}$, so that $K$ can be calculated as $M\omega_{\rm Q}^2/4=231{\rm eV}/{\rm {\AA}}^2$. The reorganization energy is $0.018{\rm eV}$ with respect to the relevant spectra, so that $\alpha$ is determined to be $3.4 {\rm eV}/{\rm \AA}$ as the equilibrated $|u_n-u_{n-1}|$ is calculated to be $0.00545{\rm \AA}$. $a_0$ is the mean distance between the cores of thiophene ring. $U$ and $V_0$ are adjustable parameters.
\end{thebibliography}
\end{document}